\documentclass[useAMS,usenatbib]{mnras}

\usepackage{amsmath} 
\usepackage{amssymb} 
\usepackage{graphicx} 
\usepackage{subfig}
\newcommand{\fixbib}[1]{}

\title[Spatial Fluctuations of the TDR]{Spatial Fluctuations of the Intergalactic Temperature-Density Relation After Hydrogen Reionization}

\author[L.C. Keating et al.]{\parbox{\textwidth}{Laura C. Keating$^{1,2,3}$\thanks{E-mail:
    lkeating@cita.utoronto.ca}, Ewald Puchwein$^{1,2}$ and Martin G. Haehnelt$^{1,2}$}\vspace{0.4cm} \\
\parbox{\textwidth}{$^1$Institute of Astronomy, University of Cambridge,
  Madingley Road, Cambridge, CB3 0HA, UK\\
$^2$Kavli Institute  for Cosmology,  University of Cambridge,
  Madingley Road, Cambridge, CB3 0HA, UK\\
  $^3$Canadian Institute for Theoretical Astrophysics, 60 St. George Street, University of Toronto, ON M5S 3H8, Canada\\}
}

\begin{document}

\date{\today}

\pagerange{\pageref{firstpage}--\pageref{lastpage}} 

\pubyear{2016}

\maketitle

\label{firstpage}

\begin{abstract}
The thermal state of the post-reionization IGM is sensitive to the timing of reionization and the nature of the ionizing sources. We have modelled here the thermal state of the IGM in cosmological radiative transfer simulations of a realistic, extended, spatially inhomogeneous hydrogen reionization process, carefully calibrated with Ly$\alpha$ forest data. We compare these with cosmological simulations run using a spatially homogeneous ionizing  background. The simulations  with  a realistic growth of ionized regions and a  realistic spread in reionization redshifts show, as expected,  significant spatial fluctuations in the temperature-density relation (TDR) of the post-reionization IGM. The most recently ionized regions are hottest and exhibit  a flatter TDR. In simulations consistent with the average TDR inferred from Ly$\alpha$ forest data, these spatial fluctuations   have  a moderate  but noticeable effect on the statistical properties of the Ly$\alpha$ opacity of the IGM  at $z \sim 4-6$. This should be taken into account in accurate measurements of the thermal properties of the IGM and the free-streaming of dark matter from Ly$\alpha$ forest data in this redshift range. The spatial variations of the TDR predicted by our simulations  are, however, smaller by about  a factor two than would be necessary to  explain the observed large spatial opacity fluctuations  on large ($\geqslant$  50 $h^{-1}$ comoving Mpc) scales at $z\gtrsim 5.5$.
\end{abstract}

\begin{keywords}
galaxies: high-redshift -- quasars: absorption lines -- intergalactic medium -- methods: numerical -- dark ages, reionization, first stars
\end{keywords}

\section{Introduction}

Reionization is expected to be an extended, inhomogeneous process \citep[e.g.,][]{madau1999,mhr2000,gnedin2000,furlanetto2004}. This appears to be reflected in the observed rapid increase of opacity fluctuations of the IGM as probed by the  Lyman-$\alpha$ (Ly$\alpha$) forest at $z>5$ \citep[e.g.,][]{becker2001, white2003, songaila2004,fan2006}. More recently, \citet{becker2015} used new measurements of the effective optical depth of the Ly$\alpha$ forest in the spectra of quasars out to $z \sim 6$ to point out that the opacity fluctuations extend to rather large scales ($\geqslant$  50  $h^{-1}$ comoving Mpc) and are larger than can be explained with the evolution of the density field alone. 

Using full radiative transfer simulations of the inhomogeneous reionization process, driven by massive stars in  high-redshift galaxies and calibrated with Ly$\alpha$ forest data,  \citet{chardin2015,chardin2017}  argued that these fluctuations of the Ly$\alpha$ opacity on large scales  may suggest a  contribution to the UV background from rare, bright sources. Alternatively, \citet{davies2016} have suggested that large fluctuations in the mean free path  may source  spatial fluctuations on large scales in a UV background from faint stellar sources,  strongly amplifying fluctuations in the density field.  Most relevant for the  work presented here, \citet{daloisio2015} considered the effect of spatial fluctuations of the temperature-density  relation (TDR) in the post-reionization Universe caused by regions ionized at different redshifts. \citet{daloisio2015} suggested  that these are responsible for the large Ly$\alpha$ opacity fluctuations on large scales due to the temperature dependence  of recombination rates. Note that the three different models make rather different predictions
for the correlation between the large scale fluctuations of the Ly$\alpha$ opacity and the space  density and brightness of ionizing sources which should -- at least in principle --  be observable (see \citealt{davies2017lae} for some recent modelling of the correlation between Ly$\alpha$ forest opacity and galaxy surface density of Ly$\alpha$ emitters).

Modelling all these effects accurately is numerically very challenging. It requires cosmological radiative transfer simulations that can properly model the temperature evolution during the inhomogeneous reionization. These simulations must also have sufficient dynamic range to resolve the sinks of ionizing radiation and  at the same time capture the  mean free path of ionizing photons, which is rapidly increasing when the individual ionized regions percolate at the end of the reionization process.  

If reionization first proceeds in an inside-out fashion \citep[e.g.,][]{furlanetto2005,choudhury2009}, then the underdense voids will be the last to reionize and should have high temperatures after overlap. However, adiabatic cooling means that the expanding voids are also the most efficient at cooling so these fluctuations may fade away quickly \citep{miraldaescude1994,uptonsanderbeck2016}. There will also be fewer recombinations in the underdense gas and the photoheating rate will therefore be lower. The effect of the hotter voids may already be unimportant for measurements of the temperature of the IGM at $z<5$ using the Ly$\alpha$ forest \citep{becker2011temp}, which is most sensitive to densities close to the mean.  The impact of spatial fluctuations of the TDR may, however, nevertheless be present in higher-redshift statistics of the Ly$\alpha$ forest, such as the aforementioned distribution of Ly$\alpha$ optical depths and the flux power spectrum. The latter is particularly important for placing constraints on the mass of warm dark matter particles \citep{viel2005,viel2013,irsic2017}.

The effects of temperature fluctuations on the post-hydrogen reionization IGM have previously been discussed in several other works, using both semi-numerical models \citep[e.g.,][]{furlanetto2009, lidz2014, daloisio2015} and radiative transfer simulations \citep[e.g.,][]{trac2008, cen2009}. We use here new full cosmological radiative transfer simulations to obtain improved estimates of the expected spatial fluctuation of the TDR  at the end of reionization.  Our emphasis will be on  models that are consistent with the temperature of the IGM at mean density inferred from Ly$\alpha$ forest data. We will perform a detailed direct  comparison of mock absorption spectra obtained from our simulations with available Ly$\alpha$ forest data.  Accurately predicting the effect  the temperature fluctuations have on  Ly$\alpha$ forest data  will eventually require careful modelling of the ionizing sources in a multi-frequency radiative transfer simulation \citep[e.g.,][]{pawlik2011}.  This is unfortunately still beyond the scope of this paper and we will instead employ mono-frequency radiative transfer simulations for a range of photon frequencies. These simulations will allow us to estimate how long  these fluctuations persist past reionization and the  length scales over which they occur. By performing  mono-frequency radiative transfer simulations for a range of photon frequencies we can also  explore the range of models consistent with the mean TDR inferred from Ly$\alpha$ forest data. 

We will further compare our simulations to estimates of the effect of spatial variations of the TDR of the IGM in the post-reionization Universe based on the hybrid approach  employed by \citet{daloisio2015}  which  combines the results of suites of optically thin simulations without radiative transfer  performed for a range of (instantaneous) reionization redshifts.  For this we will perform our own suite of optically thin simulations that are instantaneously reionized at a range of redshifts. These are then combined with a  reionization history taken from one of our radiative transfer runs.

The paper is structured as follows: in Section 2, we discuss the cosmological hydrodynamic and radiative transfer simulations we use to model the reionization of hydrogen. In Section 3, we discuss the temperature of the IGM after reionization has ended. In Section 4, we compare our simulations to several Ly$\alpha$ forest probes of the high-redshift IGM and discuss the effect of the spatially varying TDR due to inhomogeneous reionization. Finally, in Section 5, we present our conclusions. We assume the cosmology $\Omega_{\rm m} = 0.308$, $\Omega_{\Lambda} = 0.692$, $\Omega_{\rm b} = 0.0482$, $h = 0.678$, $\sigma_8 = 0.829$ and $n_s = 0.961$, consistent with the \citet{planck2015params} results.

\section{Modelling Reionization with Optically Thin and Radiative Transfer Simulations}

\begin{table*}
\centering
\begin{tabular}{c|c|c|c|c|c|c|c}
Name & $E_{\gamma}$ (eV) & Box Size (cMpc $h^{-1}$)  & $\tau$ & $z_{\textrm{reion}}$& $z_{0.5}$ &$\Delta z$ & ($a$, $z_1$, $\alpha_1$, $z_2$, $\alpha_2$)\\
\hline
RT  & 18.4 & 20 & 0.078 & 6.09 & 8.99 & 7.72 & (1.05, 7.5, 0.5, 6.0, 1.5)\\
RT hot & 23.8 & 20 & 0.078 & 6.15 & 9.16 & 7.59 & (1.1, 8.0, 0.5, 6.0, 1.5)\\
RT 40 Mpc $h^{-1}$ & 18.4 & 40 & 0.078 & 5.99 & 9.29 & 7.97 & (0.9, 7.5, 0.5, 6.0, 1.5)\\
RT fast & 18.4 & 20 & 0.063 & 5.80 & 7.59 & 5.48 & (1.05, 9.0, 2.0, 6.5, -0.5, 6.0, 1.5)\\
\end{tabular}
\caption{Summary of the radiative transfer simulations  presented in this paper. The columns show the photon energy $E_{\gamma}$, the box size, the optical depth to reionization $\tau$, the redshift of reionization  $z_{\textrm{reion}}$ (defined as the redshift where the volume-weighted mean \ion{H}{i} neutral fraction falls below $f_{\textnormal{\scriptsize{\ion{H}{i}}}} = 0.001$), the redshift at which $f_{\textnormal{\scriptsize{\ion{H}{i}}}} = 0.5$ and the extent of reionization $\Delta z$ (defined as the difference between the redshift where $f_{\textnormal{\scriptsize{\ion{H}{i}}}} = 0.9$ and $f_{\textnormal{\scriptsize{\ion{H}{i}}}} = 0.001$). The final column shows the parameters used to modify the emissivity of the \citet{haardtmadau2012} model, using equation \ref{eqn:emissivity}. Note that the RT fast run contains an additional power-law term, so the parameters shown here correspond to the terms ($a$, $z_1$, $\alpha_1$, $z_2$, $\alpha_2$, $z_3$, $\alpha_3$). All of these simulations are run on a grid with 512$^{3}$ cells.}
\label{table:simulations}
\end{table*}

As discussed in the introduction, we will model the effect of inhomogeneous reionization on spatial fluctuations of the intergalactic TDR with a suite of mono-frequency cosmological radiative transfer simulations, as well as with a hybrid approach based on a suite of optically thin simulations with a range of reionization redshifts.  The radiative transfer simulations are performed in post-processing. They use the density fields taken from snapshots of the hydrodynamic simulation which was run with a homogeneous UV background.  We use the same rate coefficients in our hydrodynamic and radiative transfer simulations. The recombination rates and collisional ionization rates are taken from \citet{huignedin1997}. The collisional excitation rates are from \citet{cen1992}. The Bremsstrahlung cooling rates are from \citet{osterbrock2006} and the Compton cooling rate is from \citet{peebles1971}.

\subsection{Optically Thin Hydrodynamical Simulations}

The cosmological hydrodynamic simulation we present here were run with the TreePM-SPH code \textsc{p-gadget3} \citep[last described in][]{springel2005gadget}. Our fiducial simulation was run in a box with side length 20 Mpc $h^{-1}$, containing $512^3$ gas elements.  We use the same initial conditions as the 20 $h^{-1}$ cMpc/$512^3$ box presented in the Sherwood Simulation Suite \citep{bolton2017}, rerun to produce a finer output of snapshots and with the rate coefficients specified above. The mass of the gas particles is $m_{\rm gas} = 8 \times 10^{5} \, h^{-1} \, M_{\odot} $ and the mass of the dark matter particles is $m_{\rm dm} = 4.2 \times 10^{6} \, h^{-1} \, M_{\odot} $. We used a gravitational softening length of 1.6  $h^{-1}$ ckpc. Our fiducial simulation uses a simplified, computationally efficient model for star formation, designed for studies of the Ly$\alpha$ forest, where gas particles with $\Delta > 1000$ (where $\Delta$ is the density in units of the mean cosmic baryon density) and a temperature $T < 10^5$ K are turned into collisionless star particles \citep{viel2004}. We assume a hydrogen mass fraction $X = 0.76$ throughout. We use the \citet{haardtmadau2012} uniform UV background, which turns on at $z=15$ and assume ionization equilibrium. This UV background has previously been shown to reproduce constraints on the temperature of the IGM at $z>3$ \citep{puchwein2015} (assuming ionization equilibrium), however it will neglect any inhomogeneous effects. We compare the thermal history and Ly$\alpha$ forest statistics computed from this simulation to those computed in our radiative transfer simulations throughout this paper (our ``Uniform UVB'' model). We also use the density fields taken from this simulation as input for our radiative transfer simulations (described below). This is to introduce some Jeans smoothing into the density field that otherwise would not be present in a simulation run without a UV background.

\subsection{Radiative Transfer  in Post-Processing}

To explore a range of inhomogeneous reionization histories, we post-process our  hydrodynamical simulation with a radiative transfer code. While this means that we  neglect the hydrodynamic response to photoheating (e.g., smoothing of the density field), it allows us to explore different reionization histories in a relatively computationally inexpensive way. We use the radiative transfer code presented and described in detail  in \citet{bauer2015}. This code solves the radiative transfer equation on a uniform Cartesian grid, using the M1 closure relation for the Eddington tensor. We have made some modifications to the treatment of the temperature. This includes the addition of adiabatic cooling due to the expansion of the Universe and adiabatic heating (cooling) due to collapse (expansion). The temperature change due to adiabatic evolution of a gas parcel is proportional to the rate of change of density, $\frac{\rm{d}\Delta}{\rm{d}t}$ \citep[e.g.][]{huignedin1997}. We approximate this term by calculating the difference in density in a cell between two subsequent snapshots of our density field. We then use this rate to calculate a heating/cooling rate due the adiabatic evolution. We have tested this method on simple models where the density is increasing/decreasing with redshift and confirmed that it produces the desired results. We find that, in our implementation and at the densities probed by the Ly$\alpha$ forest, this term is almost always subdominant to the photo-heating, or to the combination of cooling due to Hubble expansion and Compton cooling. 

Our fiducial simulation has a boxsize of 20 cMpc $h^{-1}$. To explore trends in the impact of temperature fluctuations with increasing volume, we also explore one run in a 40 cMpc $h^{-1}$ box. We recognise that these volumes are small in the context of modelling reionization \citep[e.g.][]{iliev2014}. However, as we wish to compare our results to absorption line studies of the IGM, we also require simulations with relatively high resolution \citep[e.g.][]{bolton2009}. As we will show later, even in these smaller volumes, our resolution is still not high enough to properly model the Ly$\alpha$ forest on the smallest scales (see Section \ref{subsec:fluxpower}). Our smaller volumes also allow us to run many different reionization histories down to redshift 4, calibrating our emissivities until we found good agreement with properties of the post-reionization IGM. The disadvantage of these small volumes is that we do not model temperature fluctuations due to reionization on scales larger than the size of our box, which may be important for modelling the distribution of effective optical depths of the Ly$\alpha$ forest at these redshifts. Previous work in this area has focused on semi-numerical methods for modelling inhomogeneous reionization in large volumes \citep{daloisio2015,davies2017lae}. The work presented here uses a very different method at higher resolution and should provide a useful contrast to what has been done before.

\begin{figure*}
\includegraphics[width=2\columnwidth]{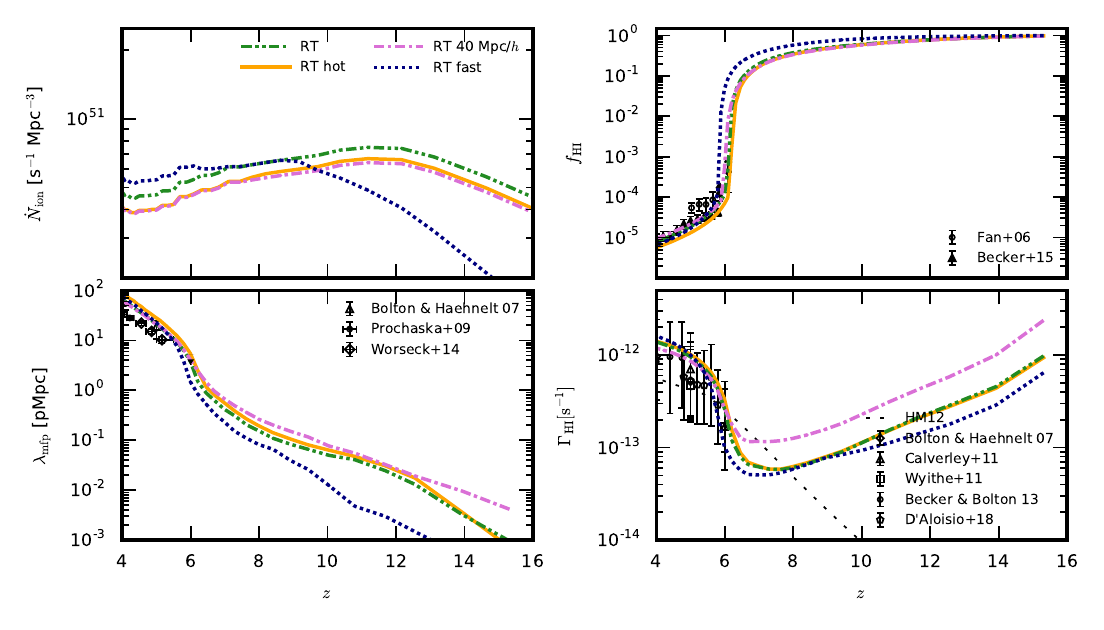}
\caption{Top left: The emissivity we assume in our radiative transfer simulations as a function of redshift.  Top right: The mean volume-weighted neutral hydrogen fraction we recover in our radiative transfer simulations. The black points are estimates of the \ion{H}{i} fraction from the opacity of the Ly$\alpha$ forest taken from \citet{fan2006} and \citet{becker2015}. Bottom left: The redshift evolution of the mean free path of a photon at 912 \AA \, measured in our simulations. Also shown are results from \citet{bolton2007gammahi}, \citet{prochaska2009}  and \citet{worseck2014}. Bottom right: The \ion{H}{i} photoionization rate as a function of redshift in the radiative transfer simulations. Shown for comparison is the photoionization rate $\Gamma_{\textnormal{\scriptsize{\ion{H}{i}}}}$ from \citet{haardtmadau2012} (black dotted line) and estimates from observations \citep[black points;][]{bolton2007gammahi,wyithe2011,calverley2011,becker2013,daloisio2018}.}
\label{history}
\end{figure*}

\begin{figure*}
\includegraphics[width=2\columnwidth]{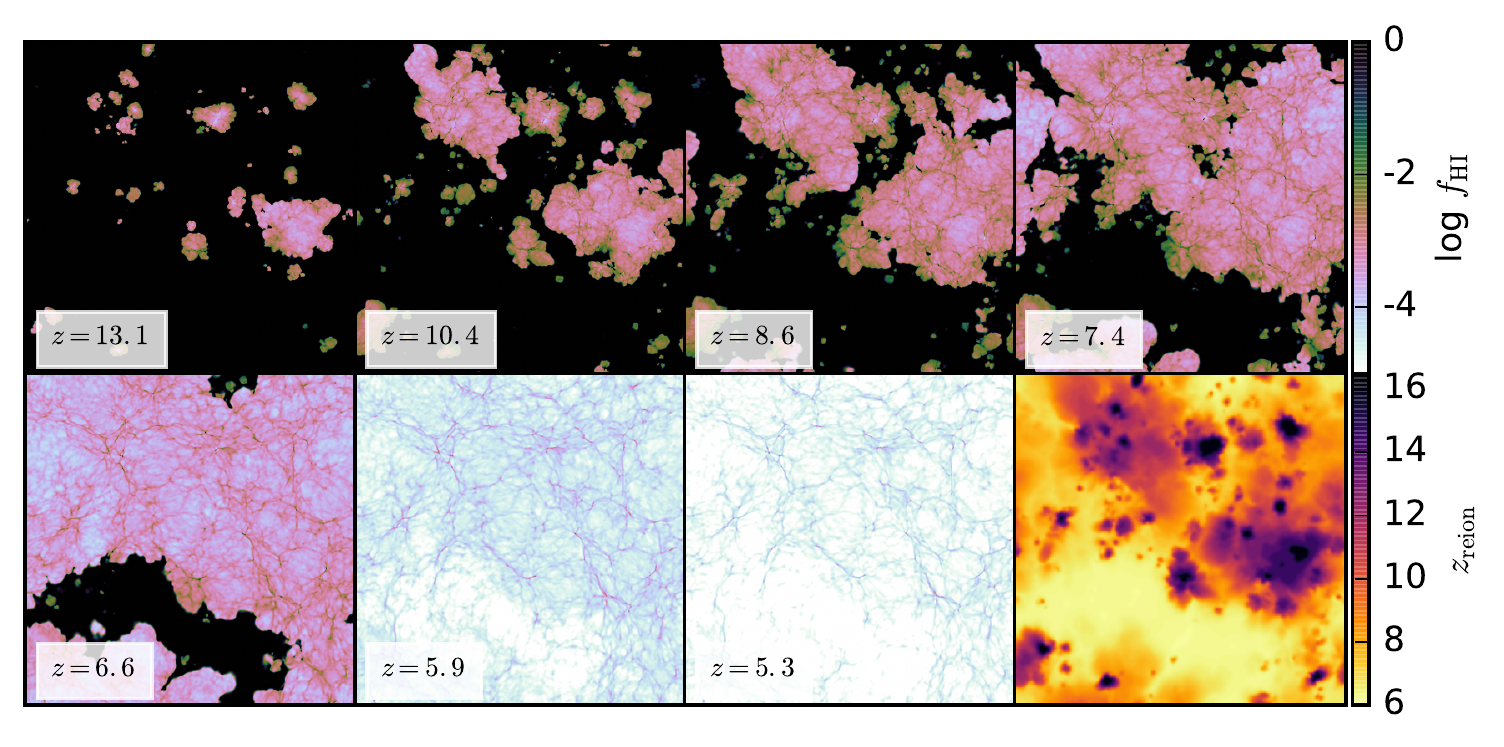}
\caption{Redshift evolution of the hydrogen neutral fraction in a slice through the RT hot simulation. The thickness of the slice is 39.1 ckpc $h^{-1}$ and the width of the slice is 20 cMpc $h^{-1}$. Bottom right: The redshift at which each cell was first ionized (defined as the redshift when the neutral hydrogen fraction first dropped below one per cent).}
\label{hfrac_zreion}
\end{figure*}

To calculate the emissivity in our simulation, we follow the method presented by \citet{chardin2015}.  We find that the emissivity required to reionize our volume is sensitive to both the resolution of the box \citep[as shown in][]{chardin2015} and also to the temperature of the gas. We assume an emissivity $\dot{N}_{\rm{ion}} = b(z)\dot{N}_{\rm{ion, HM12}}$ with $b(z)$ given by
\begin{equation} 
b(z) =
    \begin{cases}
      a & \text{if} \, \, z > z_1,\\
      a \left(\frac{z}{z_1}\right)^{\alpha_1} & \text{if} \,\, z_2 < z \leq z_1,\\
      a \left(\frac{z_2}{z_1}\right)^{\alpha_1}\left(\frac{z}{z_2}\right)^{\alpha_2} & \text{if} \,\, z \leq z_2,\\
    \end{cases}
\label{eqn:emissivity}
\end{equation}
where $a$, $z_1$, $z_2$, $\alpha_1$ and $\alpha_2$ are constants and ${N}_{\rm{ion, HM12}}$ is the emissivity from \citet{haardtmadau2012} integrated over all frequencies. We also include one run where we further lower the emissivity above $z=9$, by adding another power-law term. The evolution of the integrated emissivities we assume are shown in the top left panel of Figure \ref{history}.  We assign emissivities to haloes proportional to their dark matter masses \citep{iliev2006b} and include the contribution of all haloes with FoF dark matter mass $M_{\rm halo} > 1.2 \times 10^8 M_{\odot}$. The halo mass function in our simulation is more complete than in \citet{chardin2015} at this mass (i.e., the turnover in our halo mass function occurs at a lower halo mass than in \citet{chardin2015}), so we have a larger number of faint sources. This is likely due to differences in the halo finders used, the different gravity solvers or a different definition of the minimum number of dark matter particles for which haloes are said to be resolved. We find that we require lower emissivities than \citet{chardin2015} to reionize a volume of comparable resolution by the same redshift. This is likely  because the gas in our box is generally hotter, resulting in a lower recombination rate. Our different star formation model will also be a contributing factor, as it  removes all dense gas (rather than assuming some star formation efficiency) and this  increases the ``escape fraction'' of our galaxies and reduces the clumping factor of the gas compared to the simulations of \citet{chardin2015}. It may also be due to the differences in the halo mass functions used to define the luminosity functions in the simulations, as discussed above.

We describe the different simulations we run in Table \ref{table:simulations} and show our resulting reionization histories in the top right panel of Figure \ref{history} and in Figure \ref{hfrac_zreion}.  As discussed in detail in \citet{chardin2015} the evolution of the ionizing emissivity is tuned so that reionization ends at $z \sim 6$ in our radiative transfer runs, in agreement with estimates of the neutral fraction of the IGM using quasar absorption lines \citep{fan2006,becker2015}. In the bottom panels of Figure \ref{history} we compare our simulation to two other probes of the ionization state of the IGM: the mean free path of a photon at 912 \AA \, (bottom left panel) and the \ion{H}{i} photoionization rate (bottom right panel). We measure the mean free path in the same way as \citet{chardin2015}. For our fiducial photon energy (discussed below), we find that our mean free path is about a factor of two higher than observations. The discrepancy is larger for the RT hot run which uses a higher photon energy. The photoionization rate we calculate is likewise too high by $z=4$. This is unsurprising, since the photoionization rate scales proportionally with the mean free path. We have reasonable agreement for measurements of the photoionization rate at $z>5$, however, where most of our analysis is focused.

\begin{figure*}
\includegraphics[width=2\columnwidth]{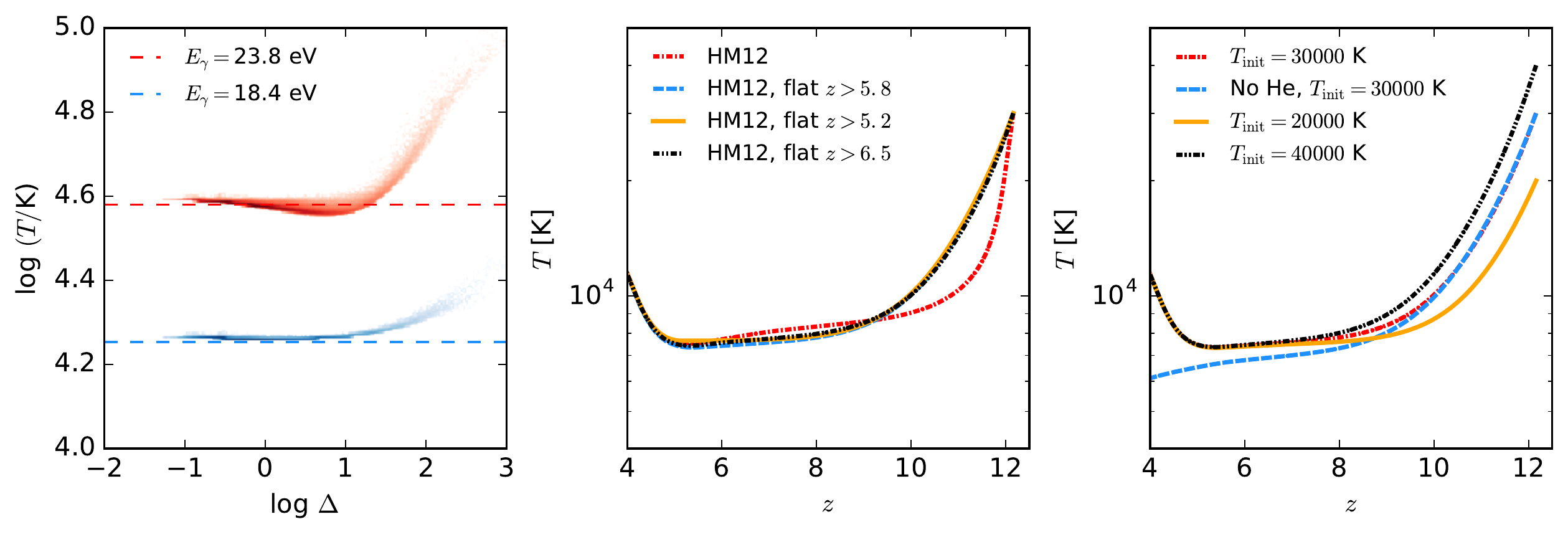}
\caption{Left: Simple test of how hot the gas in the radiative transfer simulation becomes if cooling is neglected. The red and blue contours represent two different photon energies. The dashed lines are the expected temperature for this energy directly after the gas is ionized. However, due to the finite velocity of the ionization front, we do not generally see temperatures this high in the radiative transfer simulations.  Middle: Temperature evolution of gas at mean density for different assumptions about the amplitude of the UV background, assuming the gas is initially ionized and has a temperature of 30000 K. This was calculated by solving for the thermal history of a gas parcel at mean cosmic baryon density, rather than in a full radiative transfer simulation. The evolution of the temperature with redshift does not depend on the choice of UV background, as long as the UV background is strong enough to keep the gas ionized. Right: Temperature evolution of gas at mean density, assuming different initial temperatures. Again, this was calculated by solving for the thermal history of a gas parcel at mean cosmic baryon density. Also shown is a run where heating/cooling due to helium is neglected.}
\label{temp_tests}
\end{figure*}

As explained in the introduction, we run the radiative transfer simulations in mono-frequency mode for two different photon energies. To calculate a single photon energy from this spectrum, we take the frequency averaged excess energy of the ionizing photons. This is given by 
\begin{equation} 
E_{\gamma} -E_{\textnormal{\scriptsize{\ion{H}{i}}}}   = \frac{\int^{\infty}_{\nu_{\textnormal{\scriptsize{\ion{H}{i}}}}} \mathrm{d}\nu \frac{4\pi J_\nu}{h\nu}\sigma_{\textnormal{\scriptsize{\ion{H}{i}}}} (\nu)(h\nu - h\nu_{\textnormal{\scriptsize{\ion{H}{i}}}})}{\int^{\infty}_{\nu_{\textnormal{\scriptsize{\ion{H}{i}}}}} \mathrm{d}\nu \frac{4\pi J_\nu}{h\nu}\sigma_{\textnormal{\scriptsize{\ion{H}{i}}}} (\nu)},
\end{equation}
where $E_{\gamma}$ is the total photon energy, $E_{\textnormal{\scriptsize{\ion{H}{i}}}}$ is the ionization energy of \ion{H}{i}, $\nu$ is the frequency, $J_\nu$ is the blackbody spectrum and $\sigma(\nu)$ is the cross section. Following \citet{pawlik2011}, we consider two cases for the photon energies. The first is the optically thick limit, which is calculated assuming $\sigma_{\textnormal{\scriptsize{\ion{H}{i}}}}(\nu) =1$. There is no frequency dependence here as all ionizing photons are assumed to be absorbed. The second case is the optically thin limit where $\sigma_{\textnormal{\scriptsize{\ion{H}{i}}}}$ is the usual photoionization cross section for \ion{H}{i}.  We assume that all our sources emit with a $T = 70000$ K blackbody spectrum. This was chosen so that the radiative transfer runs with photon energies corresponding to the optically thin limit provided a good match for temperatures inferred from Ly$\alpha$ forest data \citep[e.g.,][]{becker2011temp}.  \citet{pawlik2011} found that the optically thin limit produced photoheating rates in better agreement with multi-frequency radiative transfer simulations in regions far from the sources when compared to the optically thin limit.  For the optically thin limit, this corresponds to a photon energy $E_{\gamma} = 18.4$ eV. For the optically thick limit we get a photon energy $E_{\gamma} = 23.8$ eV. We also calculate the average photoionization cross section corresponding to our spectrum, giving $\sigma_{\textnormal{\scriptsize{\ion{H}{i}}}} = 2.3 \times 10^{-18} \, {\rm cm}^{2}$.

For a given photon energy $E_{\gamma}$, the temperature change $\Delta T$ of the gas as it is ionized should be
\begin{equation} 
E_{\gamma} - E_{\textnormal{\scriptsize{\ion{H}{i}}}} \approx \frac{3 k_{\rm b}}{2} \frac{\Delta T n_{\rm{tot}}}{n_{\rm{H}}}
\end{equation}
Here,  $E_{\textnormal{\scriptsize{\ion{H}{i}}}}$ is the ionization energy of hydrogen, $k_{\rm b}$ is the Boltzmann constant, $n_{\rm{H}}$ is the hydrogen number density and $n_{\rm{tot}}$ is the total number of particles. We take $n_{\rm{tot}} = (2 + Y/4X) n_{\rm{H}}$ which corresponds to ionized hydrogen and neutral helium,  where $X = 0.76$ is the hydrogen mass fraction and $Y = 1 -X$ is the mass fraction of helium. Then, for the photon energy $E_{\gamma} = 23.8$ eV ($\Delta E = 10.2$ eV), this corresponds to $\Delta T \approx  38000$ K. For the photon energy $E_{\gamma} = 18.4$ eV ($\Delta E = 4.8$ eV), this corresponds to $\Delta T \approx  18000$ K. For a test simulation where we ionize the gas instantaneously and neglect any cooling, we indeed recover this temperature. This is shown in the left panel of Figure \ref{temp_tests} for two different photon energies, representing the optically thin and thick cases for the spectrum we use throughout this work. The gas was reionized instantaneously by injecting photons into every cell. The majority of our gas has temperatures that agree well with the analytic prediction (dashed line). Gas towards higher densities has temperatures that lie above the line, likely due to additional photoheating following recombinations. In practice, however, an ionization front will move at a finite speed through the IGM and the gas will cool behind it. The peak temperatures achieved in the radiative transfer simulations are therefore usually less than the temperature changes quoted above.

\subsection{Hybrid Model}
\label{subsec:hybrid}

In order to compare our radiative transfer simulations to the hybrid approach employed by  \citet{daloisio2015} we have run a suite of simulations that are reionized ``by hand'' at a range of redshifts. We construct this suite  using eight outputs from a simulation run without a UV background. These snapshots are spaced 80 Myr apart, from $z=12.2$ to $5.9$. For each snapshot, we reinitialise  the temperature and ionization state of each gas particle. We assume the temperature of recently ionized gas is 30000 K \citep[as was assumed in][]{daloisio2015}. If the temperature of the gas particle is less than this, we set the initial temperature of the gas to 30000 K. If the temperature of the particle is greater than this, we keep the temperature of the gas particle from the hydrodynamic simulation (to account for the hotter shock heated gas). The initial ionization state of the gas is computed assuming ionization equilibrium assuming a modified version of the \citet{haardtmadau2012} UV background (described below). The simulations are then restarted from the altered snapshots and run down to $z=4$. 

We run these restarted simulations with a modified version of the \citet{haardtmadau2012} UV background, which has an increased amplitude above $z=6$ to ensure that the gas remains ionized (with the standard \citet{haardtmadau2012} background the hydrogen recombines again). The middle panel of Figure \ref{temp_tests} shows the effect of changing the UV background. We tested this using a code which solves for the temperature evolution with redshift of a gas element at mean density. If we use the standard \citet{haardtmadau2012} background, we find that this is too low at high redshift to keep the gas ionized. The partially neutral gas can then cool rapidly via collisional ionization and excitation in addition to Compton cooling (which is what we see for using \citet{haardtmadau2012}, red dash-dotted line).  As long as the gas remains ionized, the results are reasonably insensitive to the amplitude of the UV background (blue dashed and black dot-dot-dashed lines, which have photoionization rates a factor of two higher/lower than the fiducial case shown by the orange solid line). At lower redshift, we find that the model run with \citet{haardtmadau2012} is slightly hotter, due to the gas being photoheated as it is reionized. We modify the photoheating rates in the same way as the photoionization rates.

\begin{figure*}
\centering
\begin{tabular}{ccc}
\subfloat{\includegraphics[height = 5.7cm]{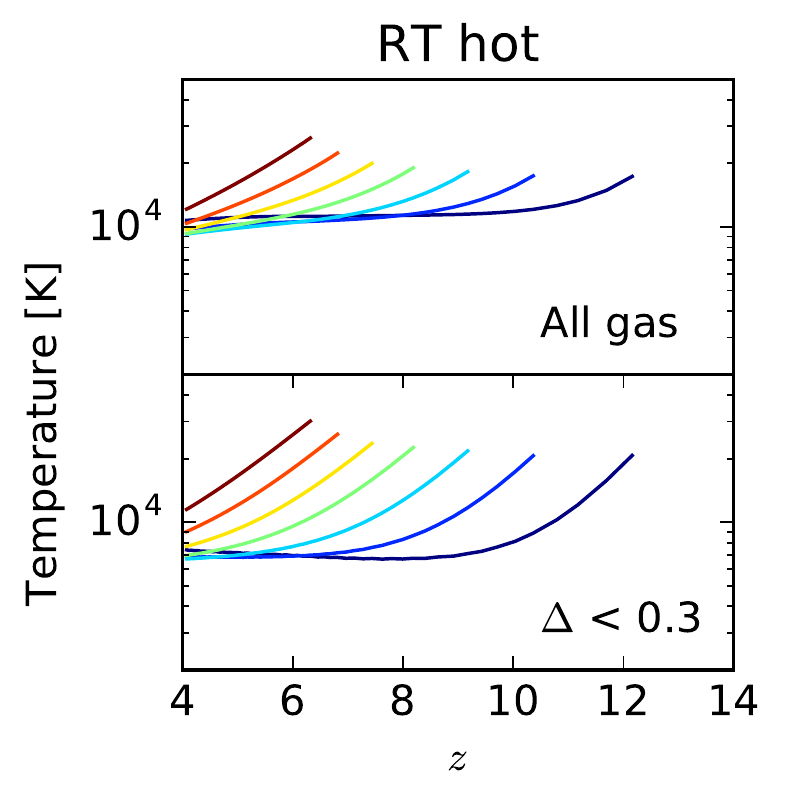}}&
\subfloat{\includegraphics[height = 5.7cm]{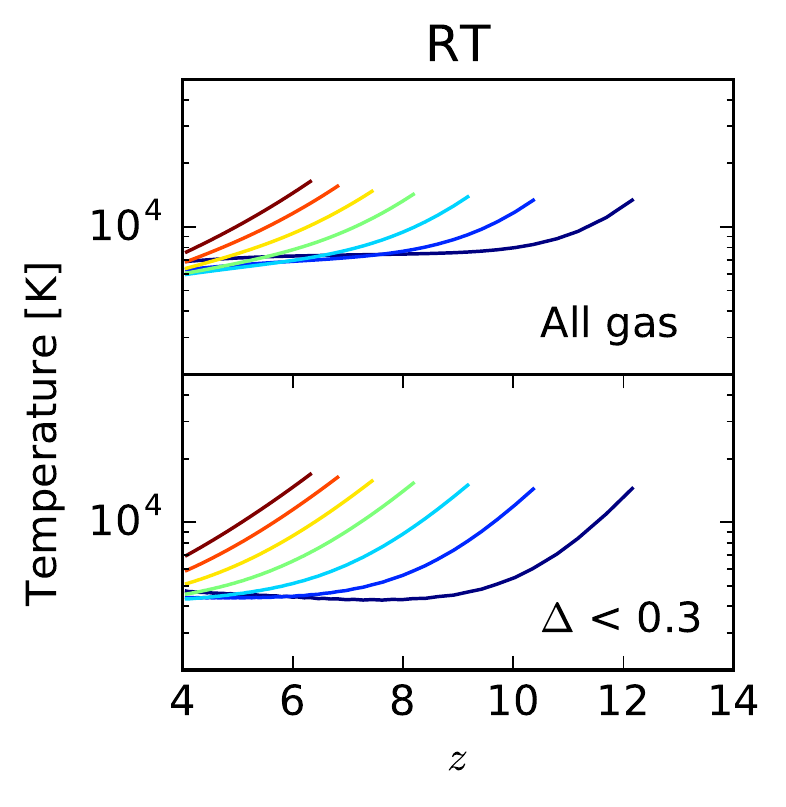}}&
\subfloat{\includegraphics[height = 5.7cm]{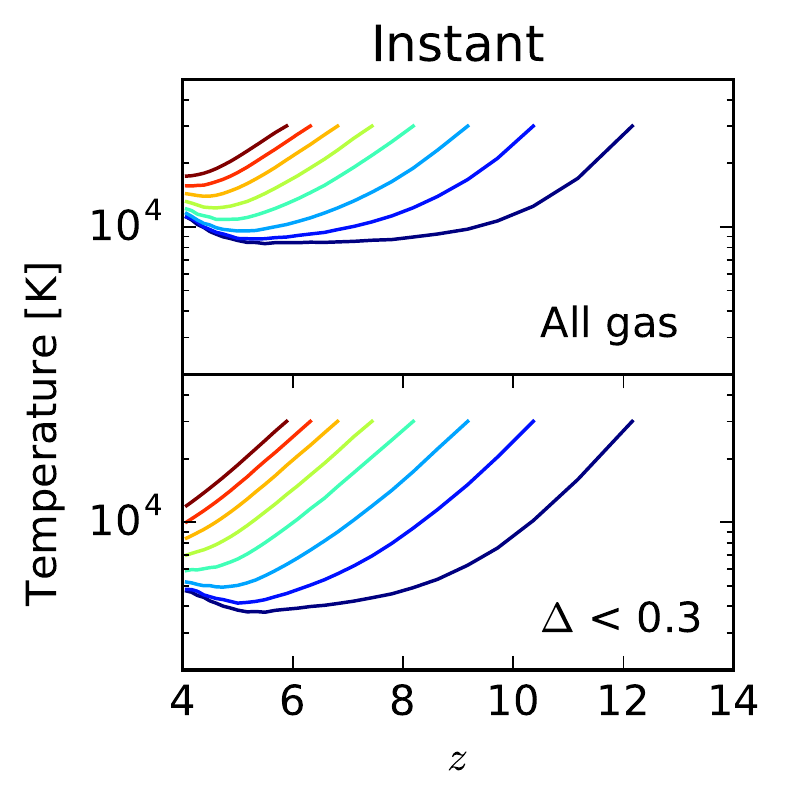}}\\
\end{tabular}
\caption{Left: Temperature evolution of cells in the radiative transfer simulation with  $E_{\gamma}$ =  23.8 eV that reionized at the same redshift for all gas (top) and gas with density $\Delta < 0.3$ (bottom). Middle: The same, but for a photon energy $E_{\gamma}$ = 18.4 eV.  Right: The same, but for the simulations that were reionized instantaneously.}
\label{temp_redshift}
\end{figure*}

We also investigate the impact of changing the temperature of the recently ionized gas (right panel of Figure \ref{temp_tests}). We increase/decrease our initial temperature by 10000 K. All the runs converge to the same temperature within about 300 Myr. We also look at the effect of neglecting helium in our thermal evolution (blue dashed line). The most noticeable difference is that the temperature begins to rise in our other models below $z \sim 5$, due to the onset of \ion{He}{ii} reionization. 

We use our grid of instantaneous reionization simulations to construct the thermal history of the IGM. This is accomplished by first calculating the reionization redshift of each cell in the RT hot simulation (which reaches temperatures similar to our hybrid model). We define the reionization redshift as the redshift when the neutral hydrogen fraction of a cell first falls below one per cent. An example of the resulting distribution of redshifts is shown in the bottom right panel of Figure \ref{hfrac_zreion}. We map the instantaneous reionization hydrodynamic simulations onto a 512$^{3}$ grid, the same resolution as our radiative transfer simulations. The hybrid model is also constructed on a 512$^{3}$ grid. For each cell of the hybrid model, we assign a temperature, density and neutral fraction taken from the instantaneous simulation that ionized at a redshift closest to the reionization redshift of that cell. This leaves us with a model that has a spatially varying temperature-density relation similar to our radiative transfer models, but with coarser time resolution (due to the finite number of instantaneous reionization simulations we have run) and which takes the hydrodynamic response of the gas due to photoheating into account.

\section{Temperature Inhomogeneities at the End of Reionization}
\label{sec:temp_dist}

\subsection{Temperature Evolution}

To compare the results from our radiative transfer and optically thin  simulations to those of \citet{daloisio2015}, we study the temperature evolution of regions of the IGM that reionized at the same redshift. In the radiative transfer simulation, we select cells that reionized at the same time and follow their thermal evolution. These redshifts are assigned by choosing the redshift of the snapshot where the neutral fraction of the cell first dropped below one per cent. This means that the time resolution of our redshifts is limited to the spacing of the snapshots, which are taken every 20 Myr. The resulting evolution of the volume-weighted mean temperature is shown in the left and middle panels of Figure \ref{temp_redshift} for radiative transfer simulations with two different photon energies. We show the temperature evolution of all the gas (top) and of gas with an density $\Delta < 0.3$ (bottom). We do not account for evolution of the density field when we select these cells. The gas associated with collapsing/expanding objects may end up in a different cell by $z \sim 4$. However, this should give an indication of the level of temperature fluctuations in the IGM that one may expect after reionization. We show here cells that reionized at redshifts in intervals of 80 Myr, to match the temporal resolution of our hybrid model. Note that in our radiative transfer simulation cells reionize in between these intervals also. In the right panel we present the temperature evolution of the eight instantaneous reionization simulations. In this case we present results averaged over the entire box, rather than a subset of cells as before.

\begin{figure*}
\includegraphics[width=2\columnwidth]{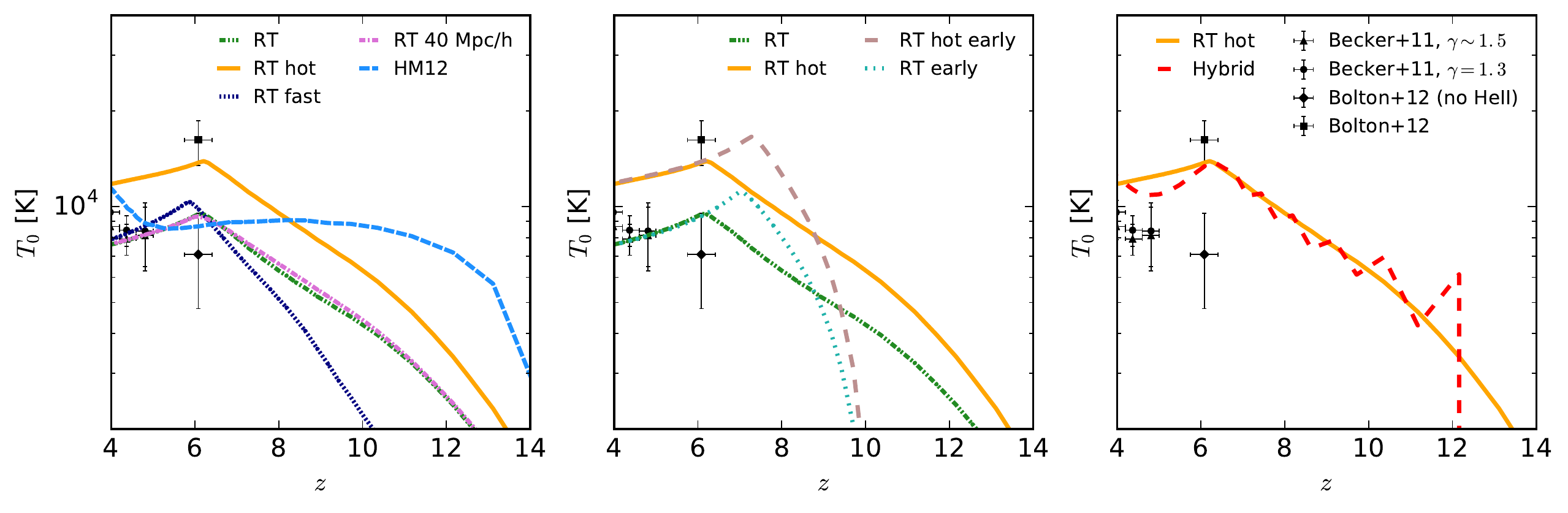}
\caption{Evolution of the volume-weighted mean temperature at mean density with redshift for the different methods of modelling the temperature evolution, different photon energies and different reionization histories. Plotted for comparison are measurements of the IGM temperature from \citet{becker2011temp} for two different values of the slope of the temperature-density relation $\gamma$, which maps the temperature measured at the characteristic density probed by the Ly$\alpha$ forest to the temperature at mean density. We also show a measurement from \citet{bolton2012} measured around a quasar near-zone and an estimate of the temperature where the likely contribution from \ion{He}{ii} heating due to the quasar is subtracted.}
\label{temp_d0}
\end{figure*}

For the radiative transfer runs, we find that the initial temperature of our ionized cells is not constant with redshift. We also do not find temperatures as high as seen in our earlier photon energy tests (left panel of Figure \ref{temp_tests}). This is partly because the cells do not ionize instantaneously, so cooling will occur in tandem with the photoheating. It is also due to the difficulty in selecting the reionization redshift in post-processing: cells that reionize just after a snapshot is taken will have had time to cool before the next output. For most of the redshift bins, the gas temperature decreases monotonically with decreasing redshift. The exception is the gas with $\Delta < 0.3$ in the highest redshift bin (at $z \sim 12$). Here, the temperature does decrease initially but then begins to increase slightly below $z=8$. This effect is likely due to how we are assigning reionization redshifts to each cell and that we are neglecting the advection of gas across cell boundaries. As expected, the low-density gas (bottom panel) cools to lower temperatures than the average of all gas due to the adiabatic expansion of the voids and the lower amount of photoheating in the low-density gas. This suggests that our simple implementation of adiabatic heating/cooling in post-processing is performing adequately. We find that the gas that reionized early (with a reionization redshift $z \sim 12$) does not cool as efficiently as the gas which ionized later. This can probably be explained by the different density distributions in the different reionization redshift bins. Gas that was ionized early on samples more biased regions of the IGM, and will have a density PDF that is skewed towards higher densities than the density PDF of the whole volume. The higher density will result in more recombinations and hence more heating by subsequent photoionizations. There may also be an additional contribution from adiabatic heating due to collapse, but we generally find that this effect is small compared to the photoheating rate.

Comparing the results of our simulations run with different photon energies, we find that the initial temperature of the gas is higher in the higher photon energy simulation as expected, but the difference is not as large as in the left panel of Figure \ref{temp_tests}. Again, this is likely due to the recently ionized gas cooling in between snapshots. The maximum temperature attained immediately after reionization is smaller at higher redshift, likely due to the increasing efficiency  of Compton cooling. The gas is also able to cool to lower temperatures in the run with $E_{\gamma} = 18.4$ eV, as the photoheating rate due to recombinations is lower in this case.

For the optically thin hydrodynamic simulations, the results are  qualitatively similar, with the gas cooling to a similar temperature to the RT run. One notable difference is the effect of including helium as already discussed in Section \ref{subsec:hybrid}. Around $z \sim 5$ the temperature of the gas begins to increase again. This upturn is present in all of our models. Note that we will not model the patchiness of \ion{He}{ii} reionization. The overall effect is to reduce the temperature contrast between our models after hydrogen reionization has ended. 

The results of our simulations  are qualitatively  similar to those presented in \citet{daloisio2015} but  suggest smaller spatial  fluctuations of the TDR. Neither the radiative transfer or hydrodynamic runs contain gas as cold ($T \sim 3000$ K for $\Delta < 0.3$ at $z =5$).  In the optically thin simulations, the scatter in temperature at $z \sim 5.5$  is more like a factor of 3 rather than the factor of 5 that \citet{daloisio2015} find. This will however be sensitive to when \ion{He}{ii} reionization is assumed to begin, but note that \citet{becker2011temp} finds that the temperature of the IGM is already beginning to increase below $z \sim 4.8$.  For the radiative transfer simulations the scatter is further reduced. 

In Figure \ref{temp_d0} we show the evolution of the temperature at mean density ($T_{0}$) with redshift for our uniform UVB, radiative transfer and hybrid models. To calculate $T_{0}$, we select cells with densities within 5 per cent of the mean and take the  volume-weighted mean of their temperatures. Also shown for comparison are measurements of the IGM temperature from \citet{becker2011temp} and \citet{bolton2012}. The measurement from \citet{bolton2012} is an estimate of the temperature around a quasar near-zone, so there is likely to be extra heating due to \ion{He}{ii} ionization close to the quasar \citep[e.g.,][]{bolton2007a,keating2015}. The effect of this extra source of heating was modelled and subtracted to estimate the temperature of the IGM at $z=6$. In Figure \ref{temp_d0}, we show the measured temperature with a contribution from \ion{He}{ii} heating and an estimate of the temperature where the likely contribution of \ion{He}{ii} heating has been subtracted. In the left panel we show results from the radiative transfer simulations presented in Table \ref{table:simulations} and our uniform UVB model. The uniform UVB model predicts gas temperatures a few times $10^{3}$ K already at $z=14$, since the \citet{haardtmadau2012} UV background turns on at $z \sim 15$. It reaches a peak at $z \sim 8$, then declines until the onset of \ion{He}{ii} reionization which begins around $z \sim 5$ when it rises again. The temperature in the radiative transfer runs rises more steeply. The gas at mean density is cooler than the uniform UVB run up until $z \sim 7$. It continues to rise until overlap occurs at $z \sim 6$ after which it cools due to Hubble expansion and Compton cooling. Increasing the energy of the photons increases the temperature of the gas somewhat. We find no trend with boxsize for the two simulations  with the same photon energy. Changing the ionization history does change the temperature evolution of the gas as expected. We have also computed the curvature of our spectra at $z = 4.8$ to compare directly to the \citet{becker2011temp}  measurements. The trend among the models is consistent with what we show here, although we do find a slightly larger difference between the models and the data (a difference of 0.1 dex in the curvature between the RT model and the \citet{becker2011temp}  measurement at $z = 4.8$). This may be due to differences in the noise treatment.

\begin{figure*}
\includegraphics[width=2\columnwidth]{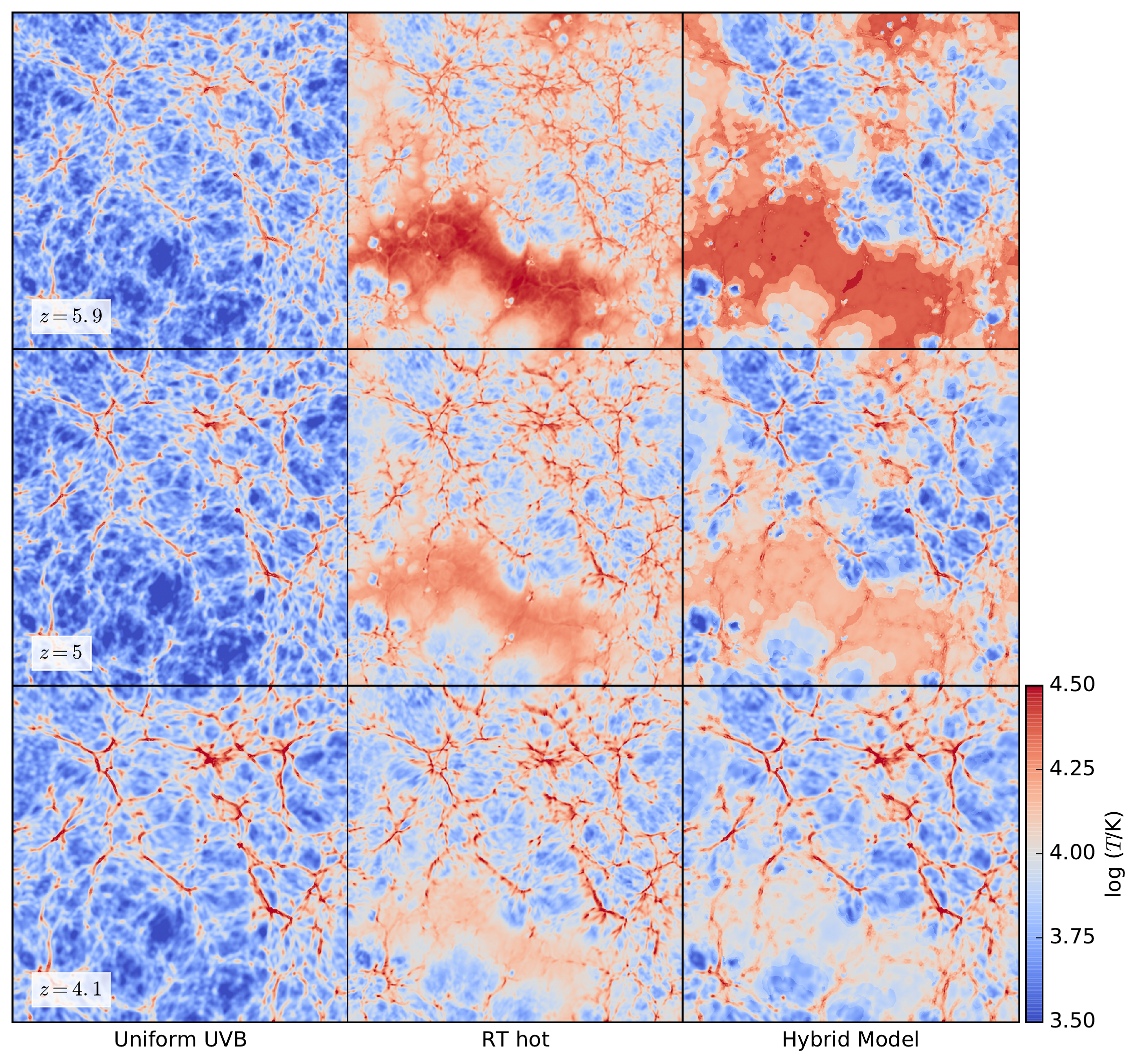}
\caption{Temperature of the gas in the same slice shown in Figure \ref{hfrac_zreion} for the three models for three post-reionization redshifts. The hottest gas in the uniform UVB simulation lies in the adiabatically heated filaments and haloes. In the radiative transfer and hybrid models, spatial fluctuations in the temperature of the IGM due to reionization are present and the recently ionized low density gas is hot.}
\label{temp_map}
\end{figure*}

In the middle panel we explore the temperature evolution of two other radiative transfer runs. We note that these runs have not been tuned to match Ly$\alpha$ forest constraints on IGM properties at lower redshifts (such as the photoionization rate) and we will therefore not discuss them elsewhere in this paper. Even so, it is still interesting to investigate their temperature evolution, which should not depend on the amplitude of the UV background. As shown in the left panel of Figure \ref{temp_d0}, we see that at $z \sim 5$, the temperature of our RT hot run is too high to be compatible with observations. We have checked  if  this could be remedied if  reionization occurred earlier, giving the gas more time to cool before the beginning of \ion{He}{ii} reionization. We explore this by looking at runs  for both photon energies we use here, where reionization is faster and finishes at $z \sim 7.5$. We find that for a faster reionization, the peak of the evolution of $T_{0}$ with redshift occurs at a higher temperature. This is because the reionization redshifts of different regions are closer together and the gas has not had time to cool significantly before reionization ends. The filling factor of hot gas will be higher and this is reflected in the average.

\begin{figure*}
\includegraphics[width=2\columnwidth]{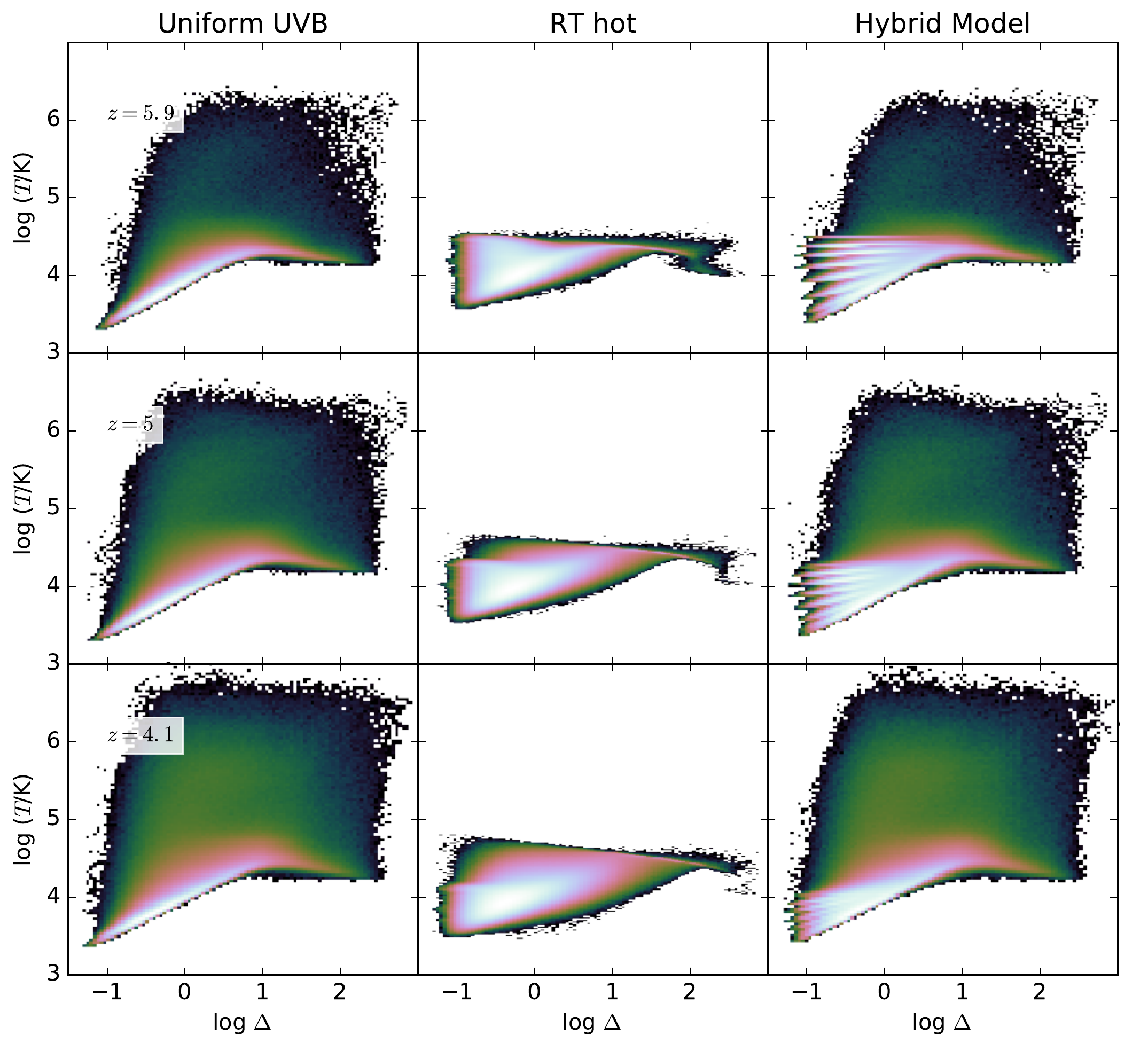}
\caption{Volume-weighted temperature-density distribution of gas in the three models, at the same redshifts as in Figure \ref{temp_map}. The radiative transfer and hybrid models both have more gas at $\log (T/\text{K}) \sim 4$ at $\log \Delta \lesssim 1$ than the uniform UVB model. No shock-heated gas ($\log (T/\text{K}) > 5$) is present in the radiative transfer run as the radiative transfer and the calculations of the thermal
evolution are performed in post-processing.}
\label{temp_dens}
\end{figure*}

Once reionization has ended in all models, the gas begins to cool and models with constant  $E_{\gamma}$ follow a common evolution for $T_{0}$ with redshift which is set by the energy injected by photoheating due to recombinations and subsequent ionizations. Therefore, even for models where reionization occurs early, it is difficult to reconcile temperature boosts as high as assumed in \citet{daloisio2015} with measurements of the IGM temperature at $z \sim 5$. This is in contrast to the results of \citet{uptonsanderbeck2016}, who find temperatures consistent with the \citet{becker2011temp} measurements assuming a linear reionization history and that gas is heated to 30000 K as it is ionized. This may be due to the different reionization history, or to the hardness of the spectrum assumed which affects how quickly the gas cools. Following hydrogen reionization the photoheating rates are expected to decrease due to a large increase in the mean-free path for ionizing photons and  the resulting change from the optically thick  to the optically  thin regime \citep{abel1999,puchwein2015, puchwein2018}. A realistic temperature evolution  may therefore fall  closer to our hot model before hydrogen reionization and closer to our fiducial model after hydrogen reionization. Investigating this properly will require multi-frequency radiative transfer simulations with realistic source spectra. 

In the right panel we compare our hybrid simulation with the radiative transfer simulation which has the closest temperature. This is the RT hot run. The temperature evolution in the  hybrid model is not entirely smooth, due to the finite number of instantaneous reionization simulations in our suite of optically thin simulations. The temperature does not increase monotonically but instead oscillates, with each peak corresponding to the redshift of one of the instantaneous simulations. Overall the evolution of $T_0$ is, however,  in good agreement with that of the corresponding radiative transfer simulation. Again, we find that this model with the higher photon energy does not agree  with the observational constraints on the temperature at lower redshift. This  suggests that models where gas receives a temperature boost as large as  $T = 30000$ K  after reionization do not agree  with the Ly$\alpha$ forest data.
   
\subsection{Temperature Fluctuations}

In Figure \ref{temp_map}, we show the projections of the IGM temperatures predicted by our three models at three different redshifts. The radiative transfer simulation  we show here is the RT hot run, which has a $T_{0}$ evolution most similar to the hybrid model. The optically thin case shows less fluctuations on large scales, as it was run with a spatially uniform UV background. The hottest gas is found in the collapsing haloes and filaments and the voids have all cooled efficiently as they expand. The radiative transfer and hybrid models both show increased temperatures ($T > 10^{4}$ K) in regions that reionized later (compare with the map of reionization redshifts in the  bottom right panel of Figure \ref{hfrac_zreion}). The regions that reionized later correspond to the low-density regions. These fluctuations persist down to $z \sim 4$ and occur on scales up to about 10 comoving Mpc, supporting the idea that they may affect statistical properties of Ly$\alpha$ forest data  in the post-reionization IGM.

Figure \ref{temp_dens} shows the volume-weighted temperature-density phase space occupied by the gas in our three models. The radiative transfer and hybrid models both contain hotter low-density gas than the uniform UVB run, due to the inhomogeneous reionization. The gas that reionized later will be hotter. This hot gas in the hybrid model looks like a discretised version of the same gas in the radiative transfer run, because  of the  finite number of optically thin simulations with different reionization redshifts we have run. Even at $z \sim 4$, there are still significant spatial fluctuations in the  temperature-density relation $T(\Delta) = T_0 \Delta^{\gamma-1}$. There is no clearly  defined overall TDR as  in the uniform UVB simulation. Indeed, in the radiative transfer run at $z \sim 6$ some of the low density gas has a temperature higher than gas at mean density. Attempting to fit a single relation to this distribution would result in a relation that flattens or inverts at low densities. This is in agreement with previous works studying the effect of hydrogen reionization on the TDR \citep{trac2008, furlanetto2009, lidz2014}. It is not clear whether the scatter in temperature should be detectable at $z \sim 4-5$. The largest scatter is found at low densities ($\log \Delta < -0.5$), but the ``characteristic density'' probed by the Ly$\alpha$ forest at those redshifts is just above the mean \citep{becker2011temp}. At lower redshifts, comparison with Ly$\alpha$ forest data is complicated by \ion{He}{ii} reionization \citep{rorai2017}.

There are several differences worth noting between the three models. Firstly, shock-heated gas (with $T > 10^{5}$ K) is visible in both the hydrodynamic runs and is missing in the radiative transfer run which does not properly model heating due to hydrodynamic effects. This is because we estimate the heating/cooling rates due to changes in $\Delta$ from the snapshots in our hydrodynamic simulation, which are spaced 40 Myr apart. However, only a small fraction of the gas is this hot. Secondly, the temperature-density relation at densities $\Delta > 1$ in our radiative transfer run is broader than the hydrodynamic runs. This seems to be a result of doing the radiative transfer in post-processing. Since the temperature is not advected with the gas, a filament or halo moving across a cell can introduce some spurious heating. The turnover in the temperature-density relation occurs at a higher density than in the hydrodynamic runs. Finally, the highest density gas in the radiative transfer simulation is colder than in the hydrodynamic runs. This corresponds to the cells where self-shielding is important, and the neutral gas can cool more efficiently. Overall, the agreement between the radiative transfer simulations and the hybrid approach using a suite of optically thin simulations is remarkably good.

\section{Comparison of Mock Absorption Spectra with Ly$\alpha$ Forest Data}

To understand the effect of these temperature inhomogeneities on observable quantities, we now test the models against statistics of the high-redshift Ly$\alpha$ forest: the distribution of effective optical depths, the flux power spectrum, the flux probability distribution function and the abundance of transmission spikes in the spectrum of ULAS J1120+0641. We construct synthetic Ly$\alpha$ absorption spectra along random sightlines through the volume, choosing the same sightlines for each of our models. These take into account peculiar velocities of the gas and the thermal broadening. The Voigt profiles are modelled as in \citet{teppergarcia2006}. In cases where we use a uniform UV background, we account for optically thick absorbers using the prescription outlined in \citet{rahmati2013}. For our hybrid model, we construct the spectra by taking sightlines though the grid we constructed from the instantaneous reionization models (as described in Section \ref{subsec:hybrid}) and computing the optical depth using the properties of the cells along the sightline.

\subsection{Effective Optical Depths}
\label{subsec:taudist}

\begin{figure*}
\includegraphics[width=2\columnwidth]{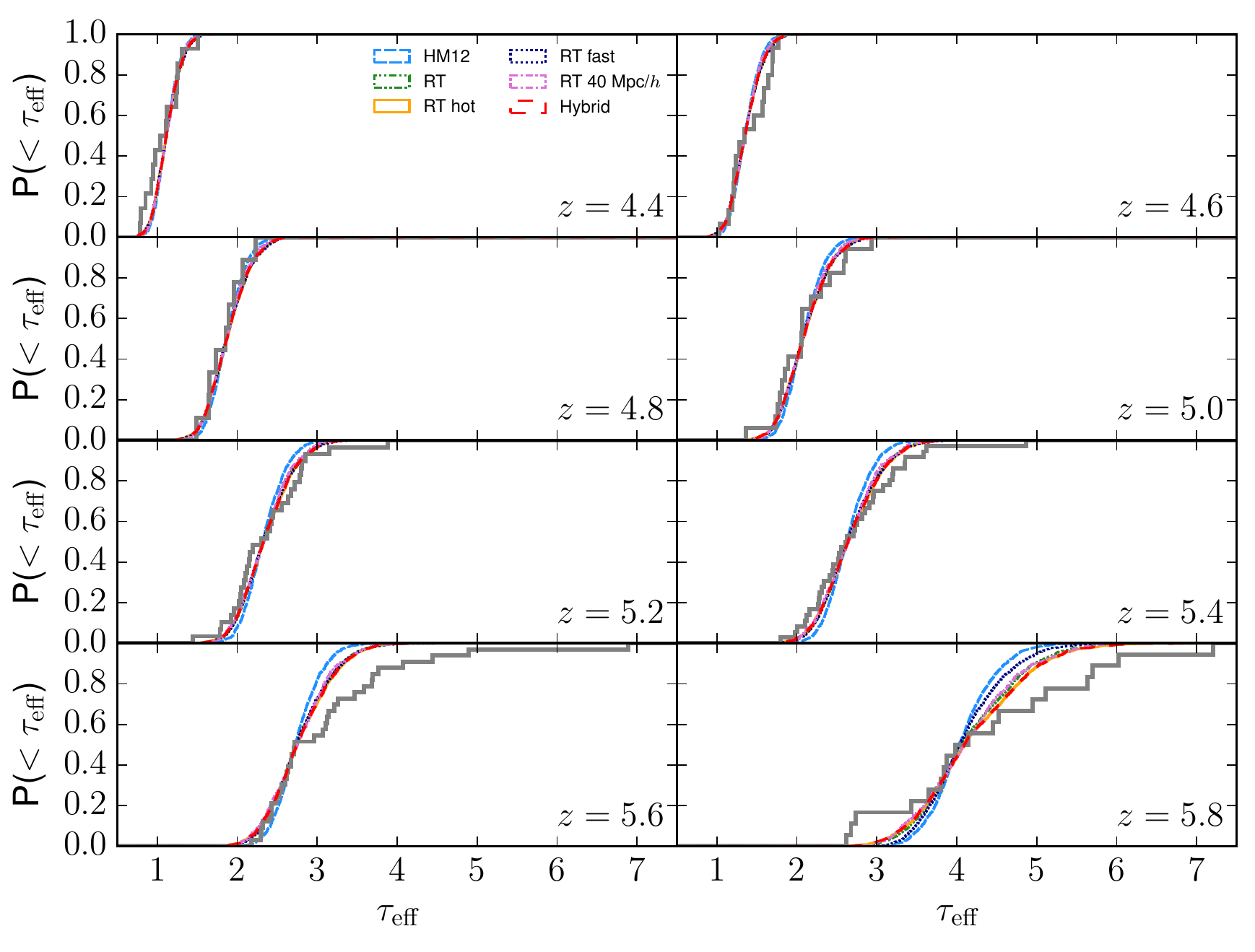}
\caption{Cumulative distribution of effective optical depths measured in 50 Mpc $h^{-1}$ chunks for the three models in eight redshift bins.  The coloured lines are distributions from our different models. Plotted in grey thick lines are the observations from \citet{becker2015} and \citet{fan2006}.}
\label{taucdf}
\end{figure*}

The effective optical depth is defined as  $\tau_{\text{eff}} = - \log \langle F \rangle$, where $\langle F \rangle$ is the mean transmitted flux. To be consistent with \citet{becker2015}, we measure this quantity along spectra with comoving length 50 Mpc $h^{-1}$. This is longer than our 20 Mpc $h^{-1}$ box, so we take 2.5 randomly selected spectra to represent one 50 Mpc $h^{-1}$ segment. We rescale the optical depth of our synthetic spectra, so that the mean flux across our whole sample matches the mean flux where the observed distribution at $P(<\tau_{\text{eff}}) = 0.5$. This rescaling ranges from factors of 0.6-3.2, with the higher values typically occurring at lower redshifts. This is likely related to the high photoionization rate in our radiative transfer simulation (Figure \ref{history}). Our effective optical depth distributions are shown in Figure \ref{taucdf} in eight redshift bins. 

As other works have found, it is not difficult to find models that agree with the observed optical depth distribution at $z \lesssim 5$ \citep{becker2015,chardin2015}. Indeed, our models fit the data reasonably well and we see little difference between the uniform UVB model and the ones that try to model reionization more accurately. Around $z=5.2-5.3$, however, we find that our radiative transfer and hybrid models begin to produce an increasingly wider distribution with increasing redshift than the uniform UVB model. It seems that this is due to the temperature fluctuations rather than any effect of the UV background. In \citet{chardin2015}, the radiative transfer simulations and the uniform UVB simulations produced nearly identical distributions in all redshift bins (after rescaling the mean flux). This is because, after overlap, there is very little fluctuation in the UV background. We similarly find a very sharply peaked distribution in our radiative transfer simulation. The temperature inhomogeneities, however, persist even after overlap has occurred, as discussed in Section \ref{sec:temp_dist}. To test this further, we constructed a set of spectra taking the temperatures of the hybrid model and calculating the \ion{H}{i} neutral fraction assuming the spatially uniform \citet{haardtmadau2012} UV background. The distribution of effective optical depths measured from these spectra is  broader than that for spectra  where temperatures are taken from the uniform UVB simulation. This suggests that temperature rather than UV fluctuations are driving this scatter in $\tau_{\text{eff}}$.

Our findings that temperature fluctuations due to patchy reionization result in a broader distribution of effective optical depths is qualitatively in agreement with the results presented by \citet{daloisio2015}.  The distribution of optical depths we recover is, however, not as broad as in  \citet{daloisio2015}. The difference may lie in exact details of the reionization history we assume, but we have chosen an extended reionization history finishing at $z \sim 6$ for our fiducial model. The difference may be due to the small volume we use in this work. Note, however, that we do not find any trends with box size, with our 20 and 40 Mpc $h^{-1}$ boxes producing nearly identical results. However, these volumes are still much smaller than in  \citet{daloisio2015}. The largest box we studied here has side length 40 Mpc $h^{-1}$ \textit{vs.} 400 Mpc $h^{-1}$ used by \citet{daloisio2015}. Indeed, they argue that their distribution becomes broader as the spatial scale of the ionized regions increases and their best fit is for regions with length larger than our simulation box size. Another difference is that they assume that more massive haloes are responsible for reionization ($1.2 \times 10^{8} \, M_{\odot}$ \textit{vs.} $2 \times 10^{9} \, M_{\odot}$). For a fixed total emissivity, using more massive haloes results in larger structures in the ionization field. This will result in variations of the temperature-density relation over larger spatial scales, and a correspondingly broader distribution of effective optical depths. The difference is unlikely to be due to our heating of the IGM due to the onset of \ion{He}{ii} reionization, as we see no difference between our radiative transfer runs (where this effect is missing) and the hybrid model (where it is included). In \citet{daloisio2015}, the authors argue that they  do not recover the largest observed optical depths in the highest redshift bins due to their semi-numerical method. This method combines a reionization history from their large 400 Mpc $h^{-1}$ volume with sightlines taken from higher resolution 12.5 Mpc $h^{-1}$ boxes, the result of which is that the reionization redshift does not correlate with density along their sightlines. In our case, this is not an issue, but nevertheless we also  do not recover these large effective optical depths due to the temperature fluctuations alone. \citet{davies2017lae} have also used a semi-numerical method to model temperature fluctuations that do correlate with the density field in a 546 Mpc $h^{-1}$ box. In their case, they are able to recover the largest effective optical depths, but note that, as discussed before, they assume higher temperatures due to reionization than in our preferred models \citep[see also][]{puchwein2018}.

Looking at the other radiative transfer runs, we find that the optical depth PDF is slightly narrower for the runs with $E_{\gamma} = 18.4$ eV. Note again that this lower photon energy was required to match the measurements of the IGM temperature (Section \ref{sec:temp_dist}). For a model that matches both the Ly$\alpha$ forest temperature constraints and has a reionization history consistent with the \citet{planck2016} measurement of the optical depth to reionization (the RT fast run), we find only a small difference from the uniform UVB model. We find no difference between the different inhomogeneous temperature models for the bins with $z \le 5.6$.  As we have discussed earlier  accurate modelling of all the relevant effects is still very challenging and will ultimately require very high-dynamic range multi-frequency radiative transfer simulations, but our simulation appear to already suggest that for models  consistent with the temperature evolution of the IGM   the spatial fluctuations of the TDR are probably about a factor two or more too small to explain the large observed  opacity fluctuations on  large  ($\geqslant$ 50  $h^{-1}$ comoving Mpc) scales at $z \gtrsim 5$. 

\begin{figure*}
\centering
\includegraphics[width=1.9\columnwidth]{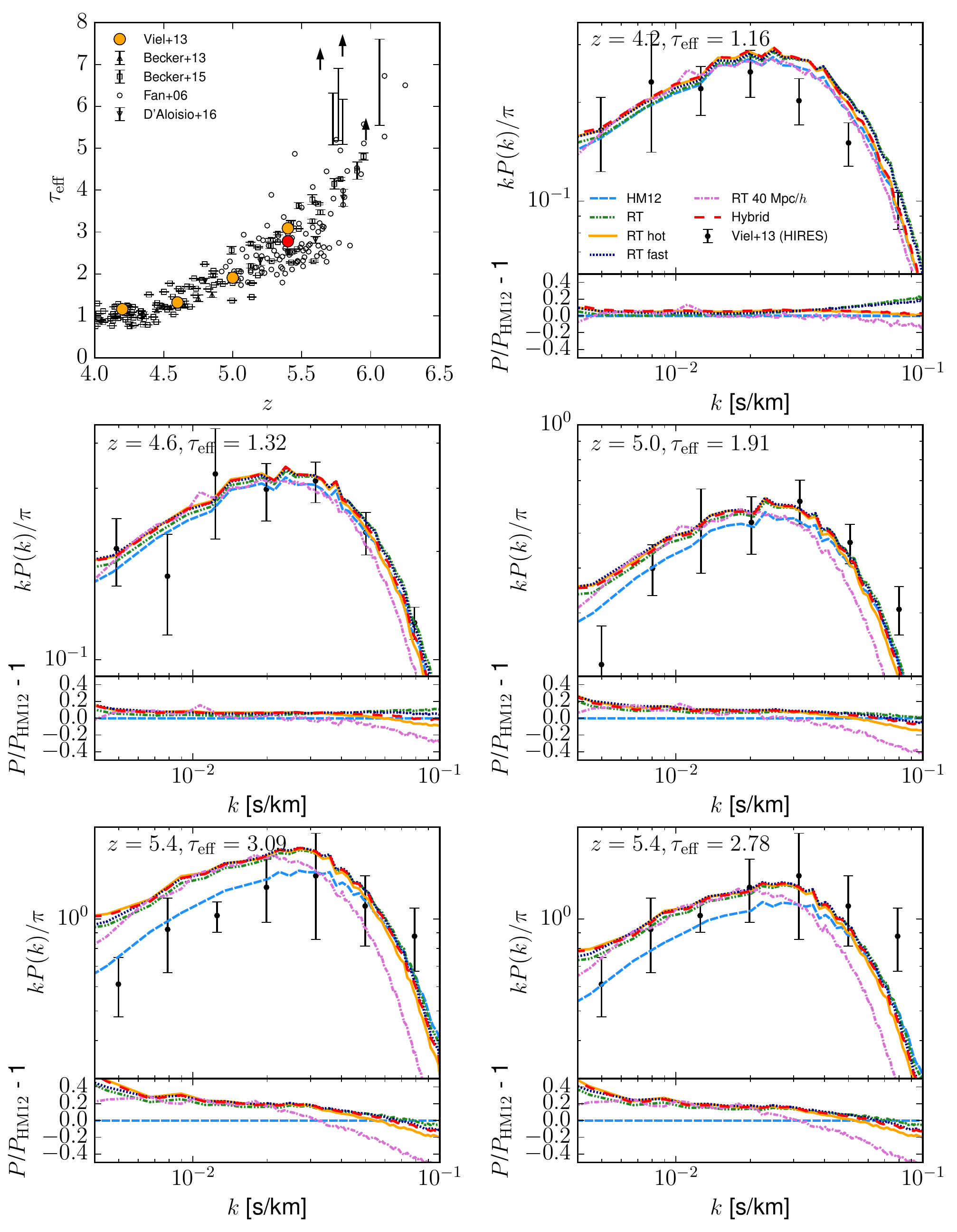}
\caption{Shown in the top left is the evolution of the effective optical depth with redshift. Shown in black are data from \citet{fan2006}, \citet{becker2013} and \citet{becker2015}. The orange circles are the best fit effective optical depths from \citet{viel2013}, to which we scale our spectra here. The red circle shows the effective optical depth at $z=5.4$ decreased by 10 per cent, to which we scale our spectra in the bottom right panel. This value is still within the 1$\sigma$ confidence limits quoted in \citet{viel2013}. The other panels show the flux power spectra for the different models in four redshift bins. The bin at  $z=5.4$ is shown twice for models rescaled to different effective optical depths. The coloured lines are distributions from our different models. Plotted in black are the observations from \citet{viel2013}. Below each panel, the power spectra relative to the  uniform UVB simulation are shown. }
\label{powerspec}
\end{figure*}

\subsection{Flux Power Spectra and Probability Distribution of Transmitted Flux}
\label{subsec:fluxpower}

\begin{figure*}
\includegraphics[width=2\columnwidth]{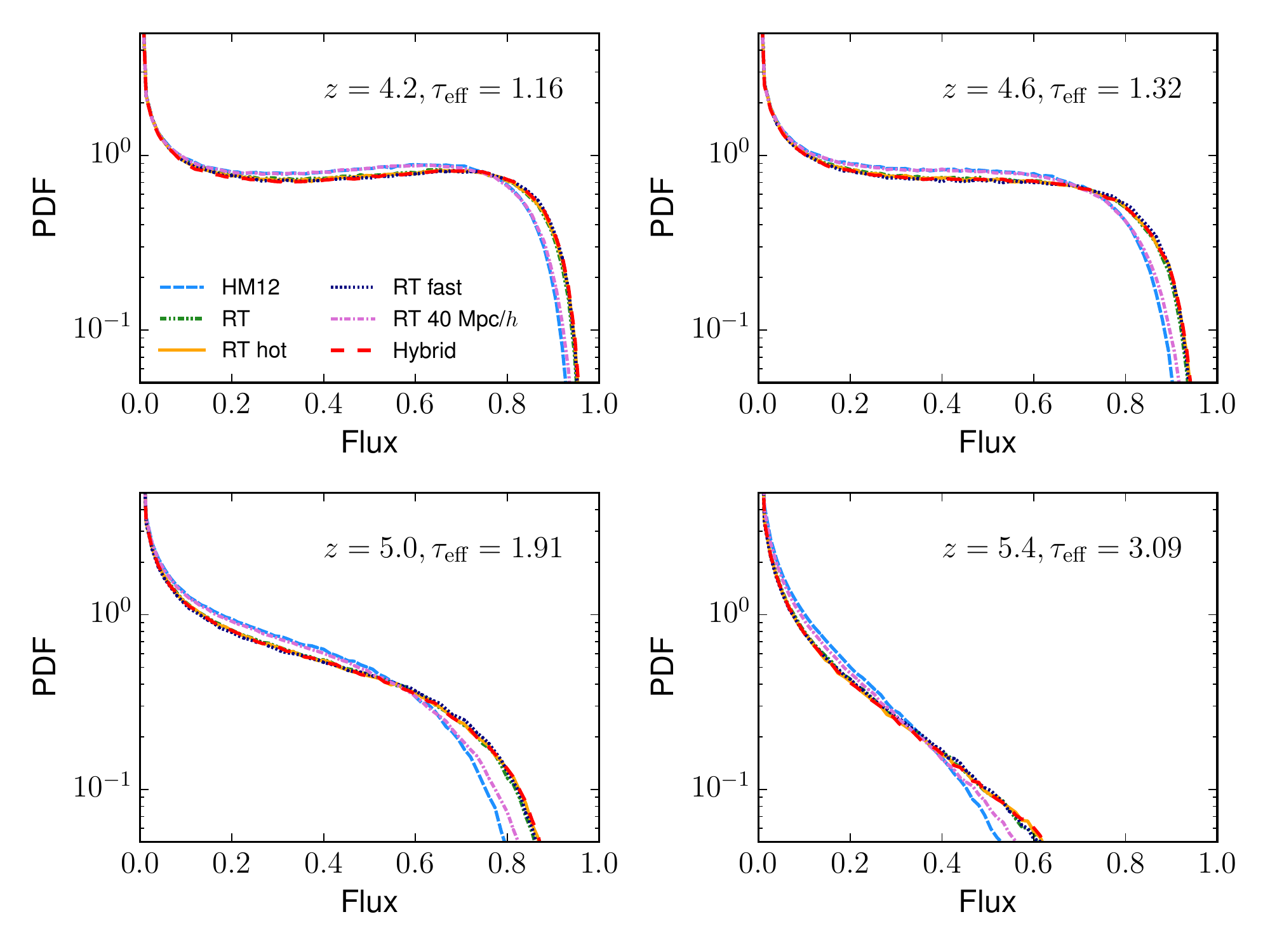}
\caption{Flux PDF in our different models in four different redshift bins, after they have been rescaled to the same mean flux and sampled onto 10 km s$^{-1}$ pixels. For simulations with the same resolution, the models with a spatially varying TDR predict more high-flux pixels.}
\label{flux_pdf}
\end{figure*}

We next investigate the effect of our different models on the high-redshift flux power spectrum. Again, at each redshift our flux is rescaled; this time to match the mean flux of the observations quoted in \citet{viel2013}. We compute the power spectrum of the fractional transmission $\delta_{\rm F} = F/\langle F \rangle -1$, where $F$ is the flux and $\langle F \rangle =  e^{-\tau_{\text{eff}}}$ is the mean flux of all sightlines at that redshift. We show the resulting power spectra in Figure \ref{powerspec} in four different redshift bins. Plotted for comparison are results for the flux power spectra from \citet{viel2013}. These power spectra are constructed from  high-resolution spectra of quasars presented in \citet{becker2007,becker2011temp,becker2011} and \citet{calverley2011}. Note that we do  not include a continuum correction as in \citet{bolton2017}. At $z = 4.2$ and 4.6 there is little difference between the models, similar to what we found for the effective optical depth distributions in Section \ref{subsec:taudist}.

We find, however,  a difference at small scales in the power spectra calculated using our 20 and 40 Mpc $h^{-1}$ boxes, suggesting that our results are not converged at this resolution \citep[see also][]{onorbe2017}. This may explain why our models are not matching the observations at the smallest scales. It is still instructive to look at the differences between the models, however. As we are doing the radiative transfer in post-processing, we are not taking into account the effect of the different pressure smoothing for different thermal histories. Any differences at large $k$ in the radiative transfer runs are therefore due to thermal broadening alone. Reassuringly, we find little difference between the RT hot run and the hybrid model (which, as it is constructed from a grid of hydrodynamic models, does account for the effect of pressure smoothing).

In the highest redshift bin, at $z=5.4$, excess power (10-20 per cent) at large scales  ($k \lesssim 7 \, h \, \text{Mpc}^{-1} = 5 \times 10^{-2}$ km$^{-1}$ s) becomes evident for the radiative transfer simulations if the optical depth in the simulated spectra matched that advocated by \citet{viel2013}. The difference  reaches about a factor of 2 increase in power at the largest scales. This is in contrast with the result of \citet{daloisio2018} who find that their temperature fluctuation model shows increased power at scales $k \lesssim 0.2 \, h \, \text{Mpc}^{-1}$ ($k \lesssim 1.5 \times 10^{-3}$ km$^{-1}$ s). This is likely due to the difference in the spatial scale of our temperature fluctuations. As we have already mentioned, the fluctuations in the simulations of \citet{daloisio2018}  occur mainly on scales larger than the box-size of our simulations. Our results are in better agreement with that of \citet{cen2009} (who simulated a 100 $h^{-1}$ Mpc volume), both in the scales at which this increased power becomes important and the amplitude of the increase. 
 
At $z=5.4$, the models with a spatially varying TDR seem to be a poorer match to the observations than the uniform UVB model. This is somewhat surprising, as in section \ref{subsec:taudist}, we found that including the inhomogeneities improved the agreement between observations and simulations at $z=5.8$. This may suggest that the  temperature inhomogeneities could play a role, but in our simulations they do not extend  to sufficiently large scales  due to the limitations of the rather small box size of our simulations.   It could also suggest that spatial fluctuations of the TDR  improving the agreement of the PDF of the opacity  at $z=5.8$ are already much less important  by $z=5.4$ (note that the uniform UVB simulation provided a reasonably good match to the effective optical depth PDF at that redshift). As shown in the upper left panel of  Figure \ref{powerspec} and as discussed before, at $z \gtrsim 5.4$ the fluctuations in the effective optical depth have begun to rise significantly. In the bottom right panel we therefore show the flux power spectrum for this redshift with a somewhat smaller effective optical depth as indicated by  the red circle in the upper left panel. This improves the agreement at large scales, but increases the discrepancy between the models and data at small scales.  Note again, however, that our simulations are not fully converged at small scales.

\begin{figure*}
\includegraphics[width=2\columnwidth]{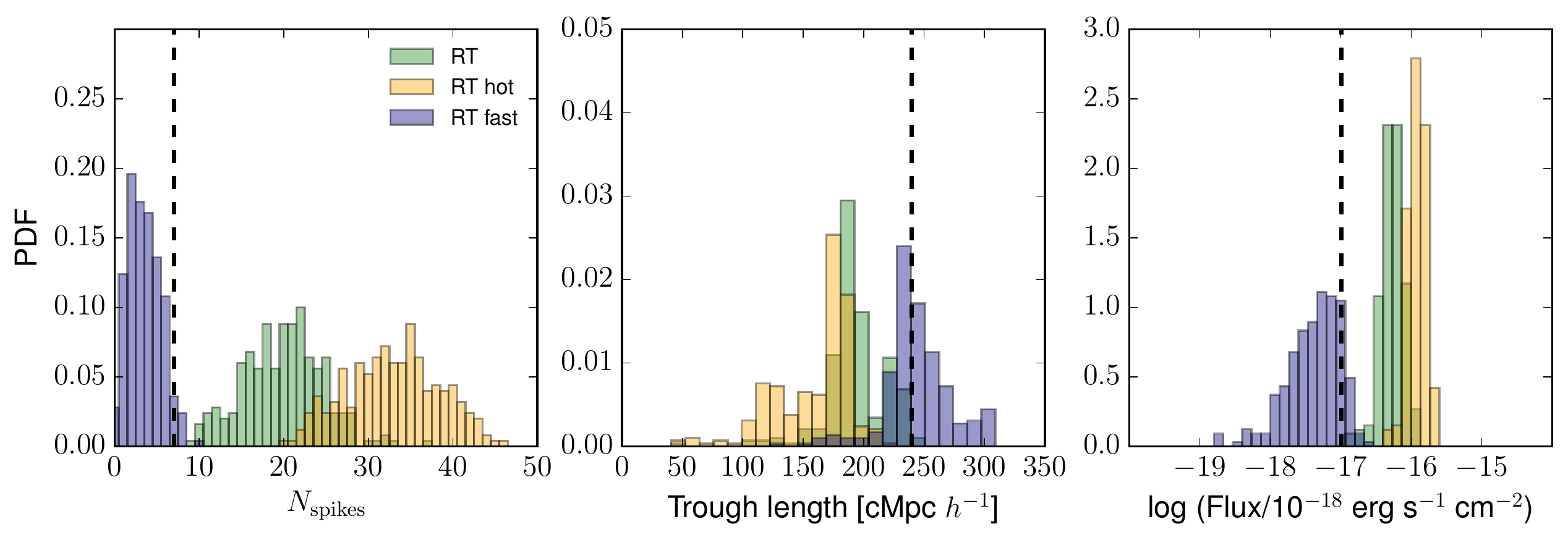}
\caption{The probability density functions of the number of spikes (left), trough lengths (middle) and total flux integrated over the detected spikes (right) measured in the RT (green), RT hot (orange) and RT fast (blue) simulations. In each panel, the black dashed line represents the measurements of each quantity taken from \citet{barnett2017}.}
\label{spikes}
\end{figure*}

Next we look at the probability density distribution of the transmitted flux (Figure \ref{flux_pdf}). Again, all the spectra are rescaled to the same mean flux. Compared with the uniform UVB simulation, the radiative transfer runs at the same resolution predict more pixels with high flux and more pixels with no flux at all. This is similar to what was found by \citet{gallerani2006} when comparing models with early and late reionization histories.  Another way of showing this difference would be to look at the distribution of dark gaps and peak heights in the spectra, as presented in \citet{gnedin2017}.

Spatial variations in the TDR of the IGM have been identified  as a source of uncertainty in Ly$\alpha$ forest constraints on warm dark matter (WDM) models \citep[e.g.][]{hui2017}. We do find that the models including temperature fluctuations due to reionization do suppress power on small scales ($k > 0.05-0.06$ km$^{-1}$ s), but this seems to be a smaller effect than in \citet{viel2013} where the suppression begins at $k \sim 0.01$ km$^{-1}$ s in their WDM models compared to the best-fit CDM model. For the RT fast run this suppression of power results in a difference of about 10 per cent at the smallest scale observed. \citet{hui2017} argue that modelling the temperature fluctuations should increase power on small scales. We do not see that here, rather seeing a suppression of power in our radiative transfer runs due to the thermal broadening of the lines. We reiterate however that our simulations are not converged at the smallest scales. Higher resolution simulations, ideally also accounting for the pressure smoothing of the gas due to photoheating, would be required to further investigate this.

\subsection{Transmission Spikes and Trough Length in the spectrum of ULAS J1120+0641}

Analysis of a VLT/X-shooter spectrum of the $z=7.1$ quasar ULAS J1120+0641 by \citet{barnett2017} showed seven transmission spikes in the Ly$\alpha$ forest in the range $5.86 < z < 6.12$, followed by a Gunn-Peterson trough of length 240 $h^{-1}$ cMpc. \citet{chardin2017spikes} recently showed that these features can be reproduced in radiative transfer simulations calibrated to match Ly$\alpha$ forest data after reionization, provided they assume an enhanced temperature in the voids of $T = 10^{4}$ K. As we already take these large-scale temperature fluctuations into account, we also investigate the occurrence of transmission spikes and troughs in our simulations. The observed trough extends over scales much larger than our 20 Mpc $h^{-1}$ simulations. However, we note that the nature of this trough is different to the deepest troughs at $z \sim 6$ in that no Ly$\beta$ flux is observed at the same redshifts. It is therefore reasonable to speculate that this trough is due to an increasingly neutral IGM, in which case our boxsize should not be a limitation.

Following \citet{chardin2017spikes}, we construct spectra by taking sightlines through six different simulation outputs. We stitch randomly selected sightlines together to produce a spectrum that covers the redshift range $5.86 < z < 7.04$. We multiplied the spectrum with a power-law $f_{\lambda} \propto \lambda^{-0.5}$ (to account for the intrinsic shape of the quasar spectrum). We sampled the spectrum onto 10 km s$^{-1}$ pixels, convolved the spectrum with a instrument profile suitable for X-shooter and added noise appropriate to this observation. We note that our results are insensitive to changing the standard deviation of the noise by a factor of 2. We searched for spikes using a Gaussian matched filter with standard deviation 15 km s$^{-1}$ and identified the regions in our spectra that had a signal-to-noise ratio greater than 5.

In Figure \ref{spikes}, we show the transmission spikes, trough lengths and total transmitted flux measured from these spectra. The trough length is defined as the distance between the redshift of the last spike and $z = 7.04$. We find that for models that have the same photoionization rate at $z \sim 6$ (the RT and RT hot runs), the different temperatures in these simulations increases the number of spikes by almost a factor of two. For the models that have the same excess photon energy but different reionization histories (the RT and RT fast runs), we again find very different distributions of spikes. This is likely driven by the difference in photoionization rate, as the temperature at $z=6$ is not very different in the two runs (see Figure \ref{temp_d0}). For the number of spikes, the length of the trough and the total integrated flux, the observations presented in \citet{barnett2017} seems to sit between the RT and RT fast runs. This may suggest that a model with this temperature is favoured, but that a photoionization rate that sits somewhere between the two models is required. This would be in the range $1-3 \times 10^{-13}$ s$^{-1}$ at $z = 6$. However, this conclusion is highly sensitive to the reionization history assumed.

To make a connection with  \citet{chardin2017spikes}, who do not model temperature fluctuations due to reionization, we also looked into the impact of changing the temperatures in the voids on these statistics by combining the photoionization rates from our radiative transfer models with the temperatures taken from our optically thin simulation. Once we assume that the voids are at a lower temperature (a few times  $T = 10^{3}$ K), the number of spikes detected drops significantly as the gas temperature drops \citep[see also][]{chardin2017spikes}. For example, with a photoionization rate taken from the RT model, the median number of spikes detected changes from 20 to 8 once the temperatures from the optically thin run are used. 

\section{Summary and Conclusions}

We have presented here radiative transfer simulations of the thermal history of the IGM during and directly after the epoch of hydrogen reionization and compared them to simulations  with a uniform ionizing UV background as well as a  hybrid approach based on a suite of optically thin simulations for a range of reionization redshifts. Our radiative transfer simulations were calibrated to match Ly$\alpha$ forest constraints on properties of the lower redshift IGM, such as the neutral fraction and photoionization rate. We explored the effects of using different energies of ionizing photons, different boxsizes and different reionization histories. We then compared these simulations to the distribution of effective optical depths and the flux power spectrum at $z>4$.

For simulations where the gas receives a temperature boost at reionization similar to that used in \citet{daloisio2015}, we find that we still cannot match the PDF of effective optical depths in the highest redshift bins. We note however that these simulations have been performed in volumes far smaller than the semi-numerical studies by \citet{daloisio2015} and \citet{davies2017lae}, who both find that temperature fluctuations can explain the fluctuations in the opacity of the  Ly$\alpha$ forest. This could be explained by effects not captured here, such as large-scale clustering of galaxies.  Using a photon energy high enough to reach such a high temperature boost also puts the simulations in tension with measurements of the temperature of the IGM at $z \lesssim 5$, and with the number of transmission spikes detected above $z = 5.86$ in the spectrum of ULAS J1120+0641 (although this will be sensitive to the assumed reionization history). For radiative transfer simulations  that match Ly$\alpha$ forest constraints on the photoionization rate and the temperature at mean density, we find that the simulations predict a significantly smaller broadening of the effective optical depth PDF. If we choose our reionization history to also fall close to the \citet{planck2016} constraint on the optical depth to reionization, we clearly fail  to match the PDF in the highest redshift bin. The spatial fluctuations of the TDR of the IGM in these models  appear to be  about a factor two too small to explain the large observed  opacity fluctuation on  large  ($\geqslant$ 50  $h^{-1}$ comoving Mpc) scales at $z \gtrsim 5.5$. Higher dynamic range and multi-frequency radiative transfer simulations will be required to answer this question more accurately. We find little difference between the results obtained from the ``hybrid'' model and the radiative transfer simulations performed in post-processing. This suggests that the hydrodynamic response of the gas does  not have  a large effect on the Ly$\alpha$ statistics we have considered.

Our radiative transfer simulations with spatial  fluctuations of the TDR of the IGM require a lower effective optical depth to match  the observed flux power spectra at the highest redshift, compared to the standard modelling with  a uniform UV background. The position of the cut-off in the power spectrum  is affected  very little  in the radiative transfer simulations. This suggests that  the effect of spatial fluctuations of the TDR of the IGM on constraints on the free-streaming of dark matter is rather small.  To make more quantitative statements, radiative transfer simulations of WDM models will be required.

\section*{Acknowledgements}
We would like to thank Jonathan Chardin for helpful discussions. We would like to thank Volker Springel for making \textsc{p-gadget3} available. We would also like to thank both Andreas Bauer and Volker Springel for making their radiative transfer code available. LCK acknowledges the support of a CITA postdoctoral fellowship, an Isaac Newton studentship, the Cambridge Trust and STFC. Support by the FP7 ERC Advanced Grant Emergence-320596 is gratefully acknowledged. EP acknowledges support from the Kavli Foundation.  This work used the DiRAC Data Analytic system at the University of Cambridge, operated by the University of Cambridge High Performance Computing Service on behalf of the STFC DiRAC HPC Facility (www.dirac.ac.uk). This equipment was funded by BIS National E-infrastructure capital grant (ST/K001590/1), STFC capital grants ST/H008861/1 and ST/H00887X/1, and STFC DiRAC Operations grant ST/K00333X/1.  DiRAC is part of the National E-Infrastructure. This research was supported by the Munich Institute for Astro- and Particle Physics (MIAPP) of the DFG cluster of excellence ``Origin and Structure of the Universe''.

\bibliographystyle{mnras} \bibliography{/data/vault/lck35/ref}

\bsp

\label{lastpage}

\end{document}